\documentclass[12pt]{article}
\usepackage{graphics,epsfig}
\usepackage{amsmath}
\usepackage{amsfonts}
\usepackage{amssymb}
\newcommand{\no}{\nonumber\\}
\newcommand{\be}{\begin{equation}}
\newcommand{\ee}{\end{equation}}
\newcommand{\ba}{\begin{eqnarray}}
\newcommand{\ea}{\end{eqnarray}}
\newcommand{\ci}[1]{\cite{#1}}
\newcommand{\bi}[1]{\bibitem{#1}}
\newcommand{\la}[1]{\label{#1}}

\newcommand{\s}{\mbox{\rm sign}}
\date{}

\begin{document}

%\begin{center}

%\baselineskip=24pt

\title{Black stars induced by matter on a brane: exact solutions}
%\baselineskip=14pt

%\vspace{1cm}

\author{A.A.Andrianov$^{a,b}$ and M.A.Kurkov$^a$\\$^a$\ \  V.A.Fock Department of Theoretical Physics, \\ Saint-Petersburg State University, 198504, St.Petersburg, Russia\\
$^b$\ \ High Energy Physics Group, Departament d'ECM and\\ Institut de Ci\`encies del Cosmos,\ 
Universitat de Barcelona,\\ Av. Diagonal 647, 08028 Barcelona, Spain}

%\preprint{...}
\maketitle
\begin{flushright}
Preprint SPbU-IP-2010-02,\quad ICCUB-10-051
\end{flushright}
\vspace{0.5cm}
\begin{abstract}
New exact asymptotically flat solutions of five-dimensional Einstein equations with horizon are found to describe  multidimensional black stars generated by matter on the brane, conceivably on high energy colliders. The five-dimensional space-time is realized as an orbifold  against reflection of a special extra-space coordinate and matter on the brane is induced by tailoring of the five-dimensional Schwarzschild-Tangherlini black hole metric.
\end{abstract}

The attractive opportunity to discover mini black holes on colliders has been within the scope of recent theoretical investigations \cite{aref}.  Black hole creation may be a consequence of strong gravity at short distances \cite{add} attainable in high energy experiments if our space is realized on a hypersurface -- three-brane in a multidimensional space-time \cite{akama0,rubak}.  One of the serious difficulties to predict these processes \cite{gregory1,gregory} is related to correct (or better exact) description of  black hole geometry when the matter universe is strictly situated on the three-dimensional brane but gravity propagates into extra space dimensions.
So far several attempts have been undertaken to find such a description \cite{brane1,casad} in which however  a control on leaking the matter into extra dimensions either was absent or only approximate at asymptotically large distances (see review in \cite{gregory,kanti}). Another problem is in appearance of delta-like singularities in matter distribution hidden under horizon for static locally stable black holes. But,
in fact, black objects  must be produced on colliders from quarks and gluons  in the high energy density evolution \cite{gingr} so that  matter always remains on the brane smoothly distributed. Therefore one expects that rather black stars  are created with matter both inside and outside an event horizon   in a finite brane-surface volume.

In this work we search for new exact solutions of five-dimensional Einstein equations with horizons  to describe multidimensional black stars generated by matter on the brane. We restrict ourselves with the construction of braneworld black stars with horizons relatively small as compared to the size of extra dimension. The  exact solutions are found for one infinite extra dimension and can be used as  guiding ones to unravel the properties of black hole objects  to be created on ultra-high energy colliders (LHC). Different ways to allocate matter are analyzed by means of tailoring two five-dimensional black holes in special coordinates and cut-and-pasting their parts on the brane under design.
Special attention is paid  to the  stress-energy tensor from the bulk viewpoint vs. effective stress-energy tensor on the brane defined with the help of Einstein-SMS equations \cite{sms}.

Let's outline how to implement  matter distribution on a brane in order to obtain  an exact solution of five-dimensional Einstein equations. We assume that the matter is localized solely on a brane and is not spread out  to the bulk.
To build a brane we search for a metric ${g}_{AB}(x,y)$ which is a bulk vacuum solution of the Einstein equations with event horizon. Let's choose a hypersurface parameterized by coordinates $x_\mu$ while $y$ is a fifth coordinate taking a constant value along this surface. The indices $A,B = (0,1,2,3),5;\, \mu,\nu = 0,1,2,3 $. Suppose that: a)\ the induced metric $g_{\mu\nu} (x,y)$ is asymptotically flat for any hypersurface $y = const$ and inherits the horizon;\quad b)\ in the chosen coordinate systems ${g}_{5B}(x,y) = 0$ and the remaining metric components provide orbifold geometry ${g}_{AB}(x,y) = {g}_{AB}(x,-y)$. These metrics are compatible with the Einstein equations.

In order to generate a brane filled by matter let's   cut a part of space while preserving the orbifold geometry,
\be {g}_{AB}(x,y) \Longrightarrow {g}_{AB}(x,|z| + a ) , \label{tail}\ee
where $a$ is an arbitrary real constant. The brane is generated at $z=0$. In this approach the metric ${g}_{AB}(x,z;a)$ remains a solution of five-dimensional Einstein equations whereas the matter is induced according to the Israel-Lanczos junction conditions \cite{Israel},
\begin{equation}
 [g_{\mu\nu}K - K_{\mu\nu}]^{+0}_{-0}= \kappa_5 \tau_{\mu\nu}, \label{israel}
\end{equation}
with $\kappa_5 = 1/ M^3_*$ and $M_*$ is a Planck scale in five dimensions.
Then the metric $g_{\mu\nu}(x,a)$ is a metric projected on the  brane $z=0$ and the induced stress-energy tensor is located on this  brane. $[K_{\mu\nu}]^{+0}_{-0} = 2K_{\mu\nu}|^{+0}$ represents the extrinsic curvature tensor discontinuity \cite{wald} defined by two limits from both sides of the brane.  We notice that the requirement of orbifolding makes a mean value of extrinsic curvature on the brane vanishing  and therefore the Regge-Teitelboim equations \cite{regteit} for passive brane dynamics \cite{akama} to be trivially satisfied \cite{binet}. Accordingly the shape of a brane is rigid.

Thus, in general, the Einstein equations in the bulk read,
\begin{equation}
^{(5)}\!G_{AB} = \kappa_5 T_{AB}, \label{Eq1}
\end{equation}
where the stress-energy tensor is taken as,
\begin{equation}
T_{AB} = \delta^{\mu}_A\delta^{\nu}_B \tau_{\mu\nu}\delta(z), \label{Eq2}
\end{equation}

In terms of the extrinsic curvature  tensor for an orbifold space-time one can reduce the five-dimensional Einstein equations up to the Shiromizu-Maeda-Sasaki ones \cite{sms} (SMS) to calculate the metric solely on the brane,
\begin{equation}
^{(4)}\!G_{\mu \nu}\equiv G_{\mu \nu} =  \kappa_5^2\Sigma_{\mu\nu} - E_{\mu\nu} \equiv \kappa_4 S_{\mu\nu},\quad \kappa_4 \equiv \frac{1}{M^2_{Pl}},\label{Eq3}
\end{equation}
where \be
\Sigma_{\mu\nu} = \frac{1}{24}\Bigl(-2\tau\tau_{\mu\nu} + 6\tau^{\sigma}_{\mu}\tau_{\sigma\nu} + g_{\mu\nu}(-3\tau^{\sigma\rho}\tau_{\sigma\rho} + \tau^2)\Bigr), \ee
where $\tau\equiv \tau_{\mu}^{\mu}. $
In order to define the conformal tensor projection $E_{\mu\nu}$ we introduce the normal vector $(n_A) = (0,0,0,0,-\sqrt{-g_{55}})$ orthogonal to the brane, its covariant counterpart $n^A = g^{AB} n_B $ and the projector on the brane $q_{AB} =g_{AB} + n_A n_B$  for the signature $(+,-,-,-,-)$. Then the above tensor is related to the conformal Weyl tensor \ci{wald} as follows,
\be
E_{\mu\nu} = ^{(5)}C^A_{BCD} n_A n^C q^B_\mu q^D_\nu;\quad E_\mu^\mu = 0 .
\ee
The SMS equations are taken in the form compatible with asymptotic flatness of brane metrics.

The effective stress-energy tensor $S_{\mu\nu}$ as seen by an observer on the brane is different from the bulk one being quadratic in $\tau_{\mu\nu}$ and is also determined by the gravitational energy flow from the bulk. However we remind that the scalar curvature does not depend on the bulk gravity flow $E_{\mu\nu}$ ,
\[^{(4)}\!R = \frac{\kappa_5^2}{12} \left(3 \tau^\mu_\nu\tau^\nu_\mu -2\tau^2\right) .\]

To prepare a suitable coordinate system
we start from the  metric describing a five-dimensional static neutral black hole \cite{tanger,emp0} in Schwarzschild coordinates
$\{t,r,\theta_1,\theta_2,\theta\}$,
\ba
g_{AB} = -\mbox{\rm diag} \left[- U(r),\, \frac{1}{U(r)},\, {R}^{2}, \, {R}^{2}
 \cos^2 \theta_{{1}} ,\,{r}^{2}
 \right], \label{i_m}
\ea
where $ U(r)= 1-{\frac {M}{{r}^{2}}}$, $M$ is related to the Schwarzschild-Tangherlini radius $M \equiv r^2_{Sch-T}$ and for brevity $R= r \cos \theta$ is introduced.
These coordinates run over the following intervals $-\infty < t < \infty$, $0 < r < \infty$, $ -\pi/2 \leq \theta_1 \leq \pi/2$, $\leq \theta_2 \leq \pi/2$, $\leq \theta \leq \pi/2$.
Let's define the Gaussian normal coordinates in respect to space-like hypersurface
 $\theta=0$. The  vector orthonormal to this hypersurface $n^{A} = \left[ 0,0,0,0,1/r \right]$ and therefore the required change of coordinates involves two variables $r = r(\rho,y)$, $\theta = \theta(\rho,y)$. These functions satisfy the geodesic equations and their solutions can be presented in the integral form,
\begin{eqnarray}
|y| &=&  \int _{\rho}^{r}\!{\frac {\mbox{\rm sign}((r-\rho))\ {x}^{2}}{\sqrt { \left( {x}^{2}-M \right)
 \left( {x}^{2}-{\rho}^{2} \right) }}}{dx}, \label{eqa5}\\
\theta &=&\rho\,\int _{0}^{y}\! \left( r \left( x,\rho \right)  \right) ^
{-2}{dx}\no&=&  \int _{\rho}^{r}\!{\frac {\s((r - \rho)y)}{\sqrt { \left( {x}^{2}-M \right)
 \left( {x}^{2}-{\rho}^{2} \right) }}}{dx}, \label{eqb5}
\end{eqnarray}
where  inside the horizon $ r  < \rho < \sqrt{M}$ and outside the horizon  $  \sqrt{M} < \rho  < r $ .

We notice that the integrals in
 \eqref{eqa5} and  \eqref{eqb5} can be expressed through the standard elliptic integrals. In particular, inside the horizon,
\ba
\!\!\!\!\!\!|y| &=& \sqrt {M} \left[ {\it
    K} \left( {\frac {\rho}{\sqrt {M}}} \right) -{\it F}
    \left( {\frac {r}{\rho}},{\frac {\rho}{\sqrt {M}}} \right)\right.\no&& \left. -{\it
    E} \left( {\frac {\rho}{\sqrt {M}}} \right) +{\it E}
    \left( {\frac {r}{\rho}},{\frac {\rho}{\sqrt {M}}} \right)  \right] ,\no
 \!\!\!\!\!\! \theta  &=& \s(y)\frac{\rho}{\sqrt{M}}\, \left[ {\it K} \left( {\frac {\rho}{\sqrt {M}}}
    \right) -{\it F} \left( {\frac {r}{\rho}},{\frac {\rho}{
    \sqrt {M}}} \right)  \right].\label{t_ellipt}
\ea
and outside the horizon,
\begin{eqnarray*}
    |y| &=& \rho  \left[ {\it K} \left( {\frac {\sqrt {M}}{\rho}}
    \right) -{\it F} \left( {\frac {\rho}{r}},{\frac {\sqrt {M}}{
    \rho}} \right) -{\it E} \left( {\frac {\sqrt {M}}{\rho}}
     \right)\right.\\ &&\left. +{\it E} \left( {\frac {\rho}{r}},{\frac {\sqrt {M}}{
    \rho}} \right)  \right]
    +{\frac {\sqrt { \left( {r}^{2}-{\rho}^{2}
    \right)  \left( {r}^{2}-M \right) }}{r}},\\
    \theta &=& sgn(y)\left[{\it K} \left( {\frac {\sqrt {M}}{\rho}} \right) -{\it
    F} \left( {\frac {\rho}{r}},{\frac {\sqrt {M}}{\rho}} \right)\right].
\end{eqnarray*}
The metric in new coordinates reads,
\ba
g_{AB} = -\mbox{\rm diag}\left[ - U(r),\, {\frac {{r}^{2}{r_{{\rho}}}^{2}}{{\rho}^{2} U(r)
 }},\, {R}^{2}
 ,\, {R}^{2}
  \cos^2 \theta_{{1}}, \, 1 \right], \label{g}
\ea
where $r = r(\rho,y), \theta = \theta(\rho,y), R = r\cos \theta, r_{{\rho}}\equiv \partial r /\partial\rho$.

\begin{figure}
%[htp]
%\epsfxsize 8cm
\includegraphics
[scale=.5]{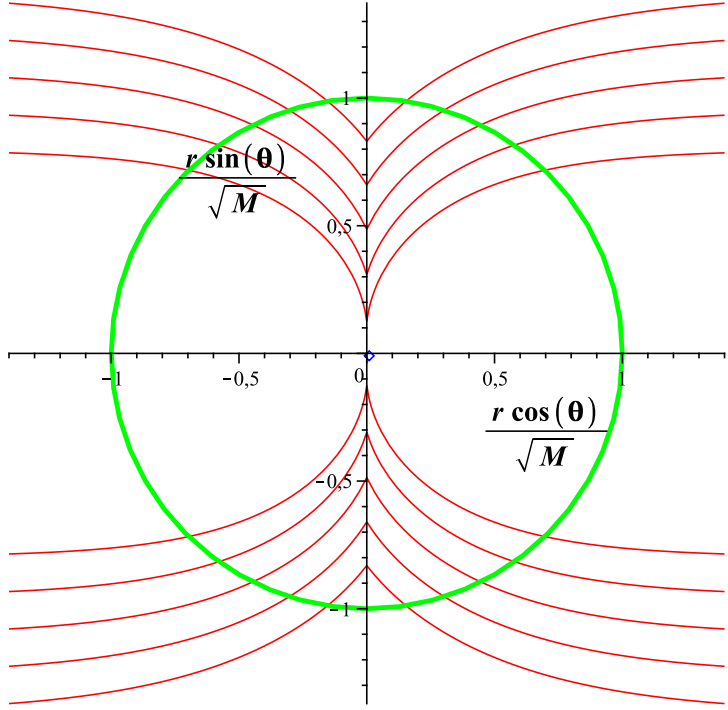}\centering
\caption{Pairs of hypersurfaces symmetric in respect to the horizontal axis   to be glued into a brane are shown by solid curves. The circle of horizon in dim = 5 is depicted long dashed.}\label{pic0}
\end{figure}

From Eq.(\ref{t_ellipt}) at  $r = 0$ one finds the minimal value $\rho_{min}$ which exists for any $y$, lies in the interval $0 < \rho_{min} < \sqrt{M}$ and grows monotonously  with increasing $|y|$.  Thus there is a lower bound for variation of $\rho$. However this lower bound in general is not equal
 $\rho_{min}$. Indeed let's analyze the limits for variation of angular variable $\theta$
 in (\ref{t_ellipt}) in the limit $r = 0$, and insert $\rho = \rho_{min}$.
 It can be shown that at fixed
  $y$ the variable $|\theta|$  monotonously decreases with increasing  $\rho$ thus one can derive $\theta_{max}$,
\begin{equation}
\theta_{max}  \equiv \mbox{sign}(y)\frac{\rho_{min}}{\sqrt{M}}\,  {\it K} \left( {\frac {\rho_{min}}{\sqrt {M}}}
    \right).
 \label{t_ellipt_zero_r_limit}
\end{equation}

Meanwhile the variable $\theta$ runs between $-\pi/2$ and $\pi/2$. Thus for large $y$ there exists a critical value  of
$\rho_{min}$ for which  $|\theta_{max}| = \pi/2$. We denote this value as  $\rho_c$. It satisfies the following equation,
\begin{equation}
\frac{\pi}{2} = \frac{\rho_{c}}{\sqrt{M}}\it K \left( \frac{\rho_{c}}{\sqrt{M}} \right);\quad \rho_{c} \simeq \mbox{0.79272} \sqrt{M}.
\end{equation}
From (\ref{t_ellipt}) at $r = 0$ one obtains the critical value
\be y_{c} \simeq 0.69868 \sqrt{M} . \la{ycrit}\ee Any hypersurface of constant $y = a$ with  $|a| < y_{c}$ contains the origin $r=0$ and therefore the stress-energy tensor $T_{AB}$ generated by \eqref{tail} reveals a point-like delta-singularity at the origin. For  $|a| > y_{c}$ the hypersurfaces intersect $~~\theta = \pi/2$ and don't pass $r = 0$, see Fig.\ref{pic0}. On such hypersurfaces minimal value of $\rho$ is larger than  $\rho_{min}$. Further on we'll study only hypersurfaces which don't intersect the singularity at $r=0$, i.e. for $|a| > y_{c}$,
to avoid delta-like singularities in matter distributions. Thus we presuppose that black stars created on colliders must possess a smooth matter density.

Let's elucidate how the horizon looks like in new coordinates.
The variable $\rho$ runs between $r$ and $\sqrt{M}$, therefore the horizon $r = \sqrt{M}$ corresponds $\rho = \sqrt{M}$.
From (\ref{eqb5})  at $r = \rho = \sqrt{M}$ it follows that on the horizon $\theta  = \frac{y}{\sqrt{M}}$, see Fig.\ref{pic0}. Taking into account the range of variation of $\theta$ one concludes that in new variables the horizon is a part of cylinder-type surface with radius  $\rho = \sqrt{M}$ and height $-\frac{\pi}{2\sqrt{M}} < y < \frac{\pi}{2\sqrt{M}}$.

Now we examine a hypersurface of constant
 $y=a$ and choose on it the following coordinates $~t, ~\rho, ~\theta_1,~\theta_2$. In accordance to
(\ref{g}) the metric induced on the hypersurface can be obtained by projection $^{(5)}\!g_{AB}\rightarrow\ ^{(4)}\!g_{\mu\nu}$ with $\mu,\nu$ ~= ~0,~1,~2,~3.

In normal Gaussian coordinates the extrinsic curvature is determined by,
\be
K_{\mu\nu} = -\frac{1}{2}\frac{\partial g_{\mu\nu}}{\partial y} \label{kmn}
\ee
Its trace,
\begin{equation}
K = {\frac {-3\,r_{{y}}r_{{\rho}}\cos \theta  -\cos
\theta  rr_{{\rho,y}}+2\,r_{{\rho}}r \sin \theta
\theta_{{y}}}{\cos  \theta  r_{{\rho}}r}}, \label{K}
\end{equation}
where $r_{{\rho}}\equiv \partial r /\partial\rho,\ r_{y}\equiv \partial r /\partial y,\ \theta_{{\rho}}\equiv \partial \theta /\partial\rho,\ r_{{\rho},y}\equiv \partial^2 r /\partial\rho\partial y$.
Let's generate a brane following the tailoring construction \eqref{tail}. Then quite remarkably the time-like component of stress-energy tensor $\tau_0^0$ is positive for a positive shift $a$ which follows from the Israel matching conditions \eqref{israel}.
The sign of $K$ is also correlated to the sign of $a$ and coincide with the sign of $\tau$ for $a>0$. Thereby positivity of $a$  is required for realization of energy conditions.

At large $\rho$ its asymptotics reads,
\begin{equation}
K = \frac{4M^2y^3}{3\rho^8}\left( 1 + O \left(\frac{y^2}{\rho^2} \right) \right),
\end{equation}
thereby confirming the asymptotic flatness of the brane. The additional evidence for the flatness is given by the asymptotics of scalar curvature,
\begin{equation}
^{(4)}\!R = \,{\frac {{4M}^{2}{y}^{2}}{{\rho}^{8}}}\left( 1 +  O \left( \frac{{y}^{2}}{\rho^2} \right) \right)
\end{equation}

On the horizon it remains finite and continuous,
\begin{equation}
K\big|_{\rho = \sqrt{M}} = \frac{-2~sgn(y)\sqrt{\frac{B}{B+1}}\cos{\frac{y}{\sqrt{M}}} + 2\sin{\frac{y}{\sqrt{M}}}}{\sqrt{M}\cos{\frac{y}{\sqrt{M}}} },
\end{equation}
where the constant $B$ represents the following limit,
\begin{equation}
B\equiv\lim _{\rho\rightarrow \sqrt {M}}{\frac {r-\rho}{\rho-\sqrt {M}}} = \frac{1}{2}\left(\cosh \left( {\frac {2y}{\sqrt {M}}} \right) - 1 \right). \label{B}
\end{equation}
In terms of this constant the values of other geometrical quantities  on the horizon can be expressed. In particular, the scalar curvature on the horizon takes the following value,
\[
^{(4)}\!R = -2\,\frac {B+1- \cos^2  y  -4\,|\sin
 y | \sqrt {1+B}\sqrt {B}\cos y }{ \left(
1+B \right)  \cos^2 y}
\]

Let's discuss the matter-density radial distribution using the Komar integral representation \cite{wald} for total mass of a black star. The total mass in 4+1 dimension is given by,
\ba
\!\!{\cal M} = \frac{3}{16\pi\kappa_5} \int\limits_{t = const} \!\!\!d^{(4)}V\,    ^{(5)}\!R_{AB}  \xi^A m^B
  \equiv \int\limits_{0}^\infty \! d R\, f_5(R), \la{mass}
\ea
where
\ba \!\!\!f_5(R) &=& -\frac{3}{8\pi\kappa_5} \int d\theta_1 d\phi\ \sqrt{|^{(4)} g|}\  K^0_0\Big|_{y=a}\no &=& -\frac{3}{2\kappa_5} \frac{r\ r_\rho}{\rho(r_\rho \cos\theta - r \theta_\rho \sin\theta)}\  K^0_0\Big|_{y=a} , \la{density} \ea
and the radial coordinate $R \equiv r(\rho,a) \cos (\rho,a)$ is chosen.
In Eq. \eqref{mass}  the Killing vector, $\xi = \partial_t$ and the vector $m$ orthonormal to the hypersurface $t = const$ are used.

A different mass distribution is seen from the brane viewpoint as being generated by the effective stress-energy tensor $S_{\mu\nu}$ in \eqref{Eq3}.
We again follow the Komar integral representation which is based on the time-like component of Ricci tensor
 $^{(4)}\!R_{0}^0$
\begin{equation}
^{(4)}\!R_{0}^0 = -{\frac {M \rho\, \left( 2\,r_{{\rho}}\,\rho\, \cos \theta  -
r \,\cos \theta +2\,\rho\,r\, \sin \theta\,
\theta_{{\rho}} \right) }{{r}^{6}\,r_{{\rho}}\,\cos \theta
}} . \label{R00}
\end{equation}
From (\ref{R00}) it can be derived that near singularity at $\theta (\rho,y) = \pi/2 $ this component is positive whereas at the infinity it is negative.
The exact calculations show that the 3-dim Komar integral \cite{wald},
\be
{\cal M}
  = \int\limits_{0}^\infty  d R\ f_4(R);\ \ f_4(R) =  \frac{8\pi r\ r_\rho\  ^{(4)}\!R^0_0}{\kappa_4 \rho(r_\rho \cos\theta - r \theta_\rho \sin\theta)}\ \Big|_{y=a} \la{mass4}
\ee on the brane vanishes which is in accordance to the tidal character \cite{casad} of black star metric $\sim 1/R^2$ at large radius.

\begin{figure}
%[htp]
%\epsfxsize 8cm
\includegraphics
[scale=.5]{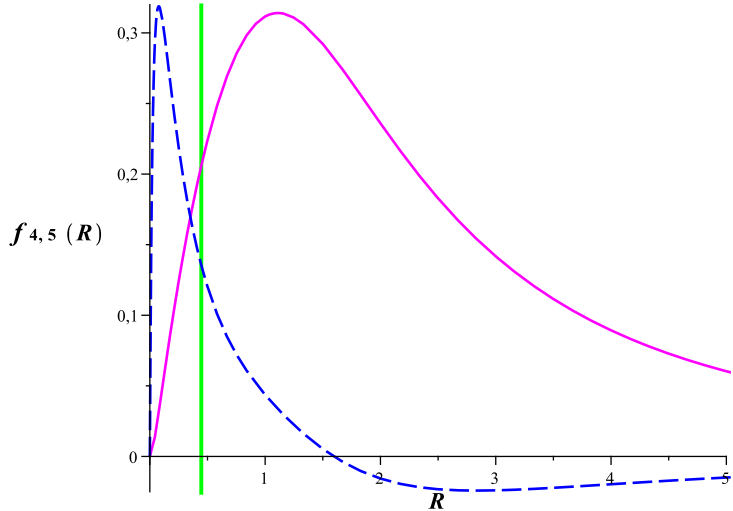}\centering
\caption{The matter-density radial distribution $f_5(R)$ on the brane with $a = 1.1, M = 1$ is presented for $\kappa_5 =1$ by a solid curve. The effective matter-density $f_4(R)$ is shown dash dotted for the value $\kappa_4 =50$ to compare with  $f_5(R)$. The horizon is indicated by long dashed line.} \label{fig1}
\end{figure}

The comparison of two radial density distributions are presented on Fig.\ref{fig1}
We see that in spite of the presence of  a singularity in
 $^{(4)}\!R^0_0$ this function is integrable in \eqref{mass4} and moreover this singularity is completely suppressed by the volume factor.

Thus we have shown that by cut-and-paste method in special Gaussian normal coordinates one can build the exact geometry of multidimensional black star with {\it horizon}, generated by  a {\it smooth} matter distribution in our universe.

In our approach,  for a given total mass, the profiles of available configurations for matter distribution are governed by the parameter $a$ which is presumably related to the collision kinematics when a black object ("black hole") is created by partons on colliders. When $a > y_c$ the very distribution does not reveal any delta-like singularity at the origin and therefore the density contribution in its small vicinity  is subdominant. For a larger $a$ the matter density happens to be more diluted approaching to the normal nuclear one and for $a > \pi\sqrt{M}/2$ the horizon disappears.
The apparent singularity at the origin $R(a) = 0$ does not lead to an acausal dynamics due to finiteness of corresponding components of metric  and affine connection. Therefore it is not a harmful naked singularity.
 To resume, one could think of the presented solution as a better approximation to describe mini black hole creation on high energy colliders than the more often used Schwarzschild-Tangherlini one.

Certainly the tailoring method used to build black objects with matter localized on branes can be generalized to the cases of charged and rotated black stars as well as black rings \cite{emp}. However more work must be done to extend the found solutions on compact extra dimensions \cite{binet,obers} and warped geometries \cite{rs}.
\section*{Acknowledgments}
We acknowledge fruitful discussions  with Prof. I.Ya.Aref'eva. This work is supported by grant RFBR 09-01-12179-ofi-m. A.A. is  partially supported by
DIUE/CUR Generalitat de Catalunya under project 2009SGR502, by the Spanish Consolider-Ingenio 2010 program
CPAN CSD2007-00042 and by MEC and FEDER under project FPA2007-66665. M.K. is supported by Dynasty Foundation stipend.

\end{document}